\documentclass{PHYEAUTH}
\usepackage{graphicx}
\usepackage{amsmath}
\usepackage{amssymb}

\begin{document}

\begin{frontmatter}

\title{
Theoretical Study of Fano Resonance in Single-Walled Carbon Nanotubes
}

\author{Junko Takahashi\thanksref{thank1}}
and
\author{Shuichi Tasaki} 

\address{Department of Applied Physics, School of Science and Engineering,
Waseda University, Tokyo, 169-8555, Japan}

\thanks[thank1]{
E-mail: junchi@ruri.waseda.jp}

\begin{abstract}
Electrical transport through single-walled carbon nanotubes (SWNTs) is investigated
by using the nearest-neighbor tight-binding model coupled with two electron reservoirs.
When the SWNT-electrode coupling is not axially symmetric, 
asymmetric resonance peaks are found in the conductance and are considered to be due to the interference 
between two transport channels. These Fano resonances are sensitive 
to the coupling with electrodes. When the coupling is axially symmetric,
no asymmetric resonance peaks are observed.  
  
\end{abstract}

\end{frontmatter}

\section{Introduction}

Fano resonance \cite{Fano}, a phenomenon that causes asymmetric conductance peaks due to 
interference between resonant and nonresonant states,
is observed in several mesoscopic systems \cite{transistor,AB ring,semiconductor,T-shaped}.
Recently,
first observations of Fano resonances are reported in crossed carbon nanotubes \cite{Kim}. 
Kim {\it et. al. } claimed that Fano resonances can not be observed in single-walled carbon
nanotubes (SWNTs) and
only appear when two nanotubes cross each other 
because the Fano resonances are attributed to the resonant scattering
by the crossed region.
Fano resonances are observed in multi-walled carbon nanotubes (MWNTs) as well, and are
considered to be due to impurities, defects or disorders \cite{Yi,Zhang}.  
Also theoretical prediction of Fano resonances for a multiply-connected carbon nanotube has been 
reported where asymmetric resonance peaks are originated from the scattering by the localized states
on the heptagonal defects \cite{Multiple Fano}.  
On the contrary, Babi\'{c} {\it et. al. } reported that, even in SWNTs
without crossed contacts, Fano resonances can be observed due to the interference 
between narrow and wide orbitals \cite{Babic}.  
    
In this report, we investigate the possibility of Fano resonances in a single armchair nanotube 
with the aid of the C$^*$-algebraic method as in our previous work on the AB-ring \cite{NESS}. 
The tight-binding Hamiltonian is adopted for 
describing the electronic states of SWNTs
and only the linear response regime is considered.   
We show that, when the SWNT-electrode coupling is axially asymmetric,
Fano-like resonances exist for an armchair SWNT.

\vspace*{-0.5cm}
\section{Electronic States of Finite SWNTs}  
\begin{figure}[t]
\begin{center}
\includegraphics[width=0.75\linewidth]{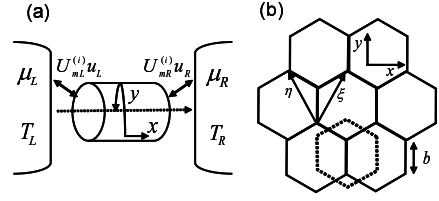}
\caption{(a) A schematic view of a finite SWNT connected to two electrodes. Tunneling strength
between each edge site of the nanotube and electrode is $U_{m\lambda}^{(i)}\cdot u _{\lambda}$. 
(b) Unit cell (dotted line) for an armchair nanotube in 2D graphite
sheet. }
\label{latticenanotube}
\end{center}
\end{figure}

We consider a system consisting of the two-dimensional electrodes and a
finite-size armchair SWNT without caps (see Fig.\ref{latticenanotube}(a)).
In this section, the electronic structure of the finite-size 
SWNT is reviewed \cite{saito}. 
The nanotube can be regarded as a rolled graphene sheet and
it is convenient to adopt the primitive lattice vectors of the graphene sheet
for specifying the atomic positions:  
\hfil \break
\indent $\vec L_{n,m}^{(1)}=n\vec \xi +m\vec \eta$,
\hfil \break 
\indent $\vec L_{n,m}^{(2)}=n\vec \xi +m\vec \eta +\vec b$, 
\hfil \break
where $n$ and $m$ are the integers, $\vec \xi =(\sqrt{3}/2,3/2) b$,
$\vec \eta =~\!\!\!(-\sqrt{3}/2,3/2)b$, 
$\vec b=(0,b)$ and $b$ is the nearest-neighbor bond length.
The armchair nanotube is constructed by rolling the graphene sheet
along the $y$-direction in Fig.\ref{latticenanotube}(b). 
The nearest-neighbor 
tight-binding Hamiltonian is used to describe the nanotube:
\begin{align}
H_{c}=\sum _{nm} &\Bigl\{-tC_{nm\sigma}^{(1)\dagger }(
C_{nm\sigma}^{(2)}+C_{n+1,m\sigma}^{(2)}+C_{n,m+1\sigma}^{(2)}) \notag
\\ &+\frac{\epsilon_{0}}{2}\sum_{i}C^{(i)\dagger }_{nm\sigma}C_{nm\sigma}^{(i)}+(\mathrm{H.c.})\Bigr\} \notag
\end{align}
where $C_{nm\sigma }^{(i)}(i=1,2)$ is the annihilation operator of the $\pi$ electron with spin 
$\sigma $ at site $\vec L_{n,m}^{(i)}$, $t$ is the nearest-neighbor hopping strength
and the on-site energy $\epsilon_{0}$ linearly depends on the gate voltage applied to the nanotube.
Throughout this report, the repeated spin index means the summation over it. 
The Coulomb interaction is neglected.
In the case of the armchair nanotube, the site $\vec L_{n,m}^{(i)}$ is identified with
$\vec L_{n+M,m+M}^{(i)}(M\in \bold N)$ and, then, the condition: 
$C^{(i)}_{nm\sigma}=C^{(i)}_{n+M,m+M\sigma}$ is imposed.
In terms of new annihilation operators:\\ 
$e_{ll'\sigma}^{(i)}=\displaystyle \frac{1}{\sqrt{M}}\sum _{n=0}^{N}\sum _{m=0}^{M-1}e^{ia_{l'}(n+2m)}S_{ln}
C_{n+m,m\sigma}^{(i)}$  \\
where $S_{nl}=\sqrt{2/(N+2)}\sin \{(n+1)\theta_{l}\}$, $a_{l'}=\pi l'/M$ and $\theta _{l}=\pi(l+1)/(N+2)$,  
the Hamiltonian of an armchair SWNT of length $(\sqrt {3}/2) bN$ becomes:
\begin{align}
H_{c}=\sum _{l=0}^{N} \sum _{l'=0}^{M-1} \Bigl[-te_{ll'\sigma }^{(1)\dagger }
&e_{ll'\sigma }^{(2)}\{1+e^{-ia_{l'}}(e^{i\theta _{l}}+e^{-i\theta _{l}})\} \notag \\ 
&+\frac{\epsilon_{0}}{2}\sum_{i}e_{ll'\sigma }^{(i)\dagger}e_{ll'\sigma }^{(i)}+(\mathrm{H.c.})\Bigr]. \notag 
\end{align}
We remark that $S_{nl}$ satisfies: 
\hfil \break
\indent $\displaystyle \sum _{l=0}^{N}S_{nl}S_{lm}=\delta _{nm}$.
\hfil \break
The Hamiltonian $H_{c}$ is , then, diagonalized by a unitary transformation: 
\hfil \break
\indent $f^{(+)}_{ll'\sigma}=P_{ll'}e^{(1)}_{ll'\sigma}+Q_{ll'}e^{(2)}_{ll'\sigma}$,
\hfil \break
\indent $f^{(-)}_{ll'\sigma}=-Q_{ll'}^{*}e_{ll'\sigma}^{(1)}+P_{ll'}^{*}e^{(2)}_{ll'\sigma}$ .
\hfil \break  
and one has
\begin{equation}
H_{c}=\sum_{l=0}^{N}\sum _{l'=0}^{M-1}\sum_{\gamma =\pm }\epsilon _{ll'}^{(\gamma)}
\; f_{ll'\sigma}^{(\gamma )\dagger }f_{ll'\sigma}^{(\gamma )} \label{hamiltonian}
\end{equation}
where $\epsilon _{ll'}^{(\gamma)}=-\gamma t\; (1+4\cos a_{l'}\cos \theta _{l}+4\cos ^{2}\theta _{l})^{1/2}$.
Note that, as $\epsilon _{ll'}^{(\gamma)}=0$ for $a_{l'}=0, \theta _{l}=2\pi/3$,
the armchair SWNT is metallic.

\vspace{-0.5cm}
\section{Steady States and Electrical Transport}
Now we consider the effect of the electrodes.
The total Hamiltonian is given by: $H=H_{el}+H_{c}+V$ where 
$H_{el}=\int dk \omega_{k\lambda }a_{k\sigma \lambda}^{\dagger }a_{k\sigma \lambda}, (\lambda =L,R)$ 
is the energy of the left or right electrode and $\omega _{k\lambda}$ is the single-electron
energy: $\omega _{kL}=\omega _{kR}+eV \propto k^{2}$ with $V$ the source-drain voltage.
The second term $H_{c}$ stands for the nanotube Hamiltonian (\ref{hamiltonian}) and
\hfil \break
\indent $\displaystyle V=\sum_{l=0}^{N}\sum_{l'=0}^{M-1}\sum _{\gamma\in \pm, \lambda\in L,R}\{h_{ll'}^{(\gamma,\lambda)}
a_{\sigma}^{(\lambda)}(u)^{\dagger}f_{ll'\sigma}^{(\gamma)}+(\mathrm{H.c.})\}$
\hfil \break
represents the tunneling between the nanotube and electrodes.
We denote $a_{\sigma}^{(\lambda)}(u)=\int dk u_{k\lambda}a_{k\sigma \lambda}$ and 
$h_{ll'}^{(\gamma,\lambda)}$ are given by:
\hfil \break
$h_{ll'}^{(\pm,L)}=\displaystyle \frac{S_{0l}}{\sqrt{2M}}\sum _{m=0}^{M-1}\Bigl(\pm U_{mL}^{(1)}
e^{-i\tau _{ll'}^{(-)}}+U_{mL}^{(2)}e^{-i\tau _{ll'}^{(+)}}\Bigr)e^{\mp i\kappa _{ll'}} $
\hfil \break
$h_{ll'}^{(\pm,R)}=\displaystyle \frac{S_{Nl}}{\sqrt{2M}}\sum _{m=0}^{M-1}\Bigl(\pm U_{mR}^{(1)}
e^{-i\tau _{ll'}^{(-)}}+U_{mR}^{(2)}e^{-i\tau _{ll'}^{(+)}}\Bigr)e^{\mp i\tilde \kappa _{ll'}}$
\hfil \break
where $\tau^{(\pm)}_{ll'}=2a_{l'}m\pm\kappa _{ll'}$, $\tilde \kappa _{ll'}^{(\pm)}=\kappa _{ll'}\pm a_{l'}N$,
$\kappa_{ll'}=~\!\!\!\frac{\delta _{ll'}}{2}$, \\$\delta _{ll'}$ is the phase of $-(1+e^{-ia_{l'}}
(e^{i\theta _{ll'}}+e^{-i\theta _{ll'}}))$ and the product $U_{m\lambda}^{(i)}\cdot u_{k\lambda}$
stands for the tunneling strength between each edge site of the nanotube and the electrode $\lambda$.

In order to calculate the steady-state current from the left to right electrodes,
we derive the asymptotic in-coming fields $\beta _{k\sigma \lambda}$ that are
solutions of the Heisenberg equation: $[H,\beta _{k\sigma \lambda}]=-\omega _{k\lambda}\beta _{k\sigma \lambda}$.
They are given by:
\begin{eqnarray}
\beta_{k\sigma \lambda}&=&a_{k\sigma \lambda}+\displaystyle \sum_{l=0}^{N}\sum_{l'=0}^{M-1}\sum_{\gamma \in \pm}\biggl\{
\dfrac{u^{*}_{k\lambda}}{\Lambda (\omega_{k\lambda})}\dfrac{\chi_{ll'}^{(\lambda)}
f_{ll'\sigma}^{(\gamma)}}{\omega_{k\lambda}-\epsilon_{ll'}^{(\gamma)}-\epsilon _{0}}
\nonumber \\ 
&&+\displaystyle \int dk'\biggl(\dfrac{u_{k'\lambda}u^{*}_{k\lambda}}{\omega_{k\lambda}-\omega_{k'\lambda}-i0}
\dfrac{\eta_{ll'}^{(\lambda)}(\omega _{k\lambda})a_{k'\sigma \lambda}}{\Lambda(\omega_{k\lambda})}
\nonumber \\
&&\mskip 50mu +\dfrac{u_{k'\bar \lambda}u^{*}_{k\lambda}}{\omega_{k\lambda}-\omega_{k'\bar \lambda}-i0}
\dfrac{g_{\lambda \bar \lambda}(\omega _{k\lambda})a_{k'\sigma \bar \lambda}}{\Lambda(\omega_{k\lambda})}\biggr)\biggr\} \label{eq:operator}
\end{eqnarray}
where $\eta _{ll'}^{(\lambda)}(\omega _{k\lambda})=\nu({\omega _{k\lambda}})
M_{\bar \lambda}(\omega _{k\lambda})+g_{\lambda \lambda}(\omega_{k\lambda})$, \\
$\chi _{ll'}^{(\lambda)}(\omega _{k\lambda})=h_{ll'}^{(\gamma,\lambda)}(1-g_{\bar \lambda \bar \lambda}
M_{\bar \lambda}(\omega _{k\lambda}))+h_{ll'}^{(\gamma,
 \bar \lambda)}
g_{\lambda \bar \lambda}M_{\bar \lambda}(\omega _{k\lambda})$,\\
$\nu(x)=|g_{LR}(x)|^2-g_{LL}(x)-g_{RR}(x)$, \\
$\Lambda (x)=1-(g_{LL}M_{L}(x)+g_{RR}M_{R}(x))-\nu(x)M_{L}(x)M_{R}(x)$,\\
$g_{\lambda \bar \lambda}(x)=\displaystyle \sum_{ll'\gamma}
\dfrac{h_{ll'}^{(\gamma,\lambda)}h_{ll'}^{(\gamma,\bar \lambda)*}}{x-\epsilon_{ll'}^{(\gamma)}-\epsilon_{0}}$, 
$M_{\lambda}(x)=\int d\tilde k \dfrac{|u_{\tilde k\lambda }|^2}{x-\omega _{\tilde k\lambda }-i0}$.
In the above, $\bar L=R, \bar R=L$. 
As in \cite{NESS}, if the system admits no bound state, a nonequilibrium steady state (NESS)
is rigorously constructed from the time-evolution. The NESS is found to satisfy Wick's theorem
and is fully specified by the 
two-point function:  
\begin{equation}
\langle \beta _{k\sigma \lambda }^{\dagger } \beta _{k'\sigma '\lambda '}\rangle =
F_{\lambda }(x)\delta (k-k'
)\delta _{\sigma \sigma '}\delta _{\lambda \lambda '}\label{eq:soukan}
\end{equation}  
where $F_{\lambda}(x)$ indicates the Fermi-Dirac distribution function of the electrode $\lambda$ with
temperature $T_{\lambda}$ and chemical potential $\mu_{\lambda}$. \\ 
The current operator is the time-derivative of
the left-electrode number operator: 
$n_{k\sigma L}=\int dk a_{k\sigma L}^{\dagger}a_{k\sigma L}$ 
and its average is
\begin{equation}
\langle \hat J \rangle=\dfrac{e}{h}\sum_{l=0}^{N}\sum_{l'=0}^{M-1}\sum_{\gamma \in \pm}
\left \{h_{ll'}^{(\gamma,L)}\langle a^{(L)}_{\sigma}(u)^{\dagger}f_{ll'\sigma}^{(\gamma)}\rangle +(\rm c.c )\right\}.
\end{equation}
By expressing $a_{\sigma}^{(L)}(u), f_{ll'\sigma}^{(\gamma)}$ with
the asymptotic field operators $\beta _{k\sigma \lambda}$ and by using (\ref{eq:soukan}), the steady-state 
current casts into the Landauer formula \cite{Landauer}:
\begin{equation}
\langle \hat J \rangle =\dfrac{e}{h} \int _{\omega_{c}}^{\infty}dx \; T(x) \times \{F_{L}(x)-F_{R}(x)\} \label{eq:denryu}
\end{equation}
where $\omega _{c}=$max$\{\omega_{kL},\omega_{kR}\}_{k=0}$.
The transmission coefficient $T(x)$ is 
\begin{equation}
T(x)=\dfrac{4\pi^2u^4D^2}{|\Lambda (x)|^2}
|g_{LR}(x)|^2 \notag
\end{equation}
where $D$ is the density of states of the electrodes and other functions are defined after (\ref{eq:operator}).
Here, we are interested in the linear response regime at $T_{L}=T_{R}=0$ and 
the variables $u_{kL}$ and $u_{kR}$ are taken to be the same and $k$-independent, i.e., $u_{kL}=u_{kR}=u$.

\vspace{-0.5cm}
\section{Fano Resonances in SWNTs}
\begin{figure}[h]
\begin{center}
\includegraphics[width=1\linewidth]{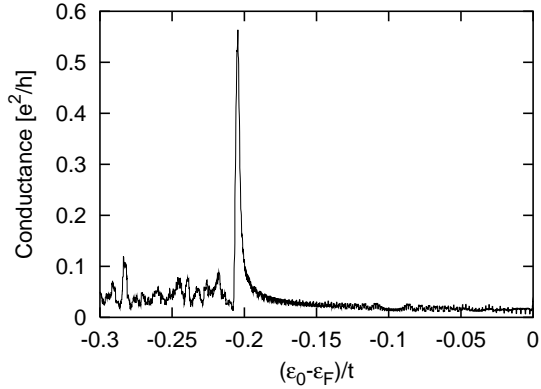}
\caption{One of the asymmetric conductance peaks in the linear response regime
as a function of the gate voltage. The parameter
is taken as $u^4D^2/t^2=1.0$ and (N,M)=(7000,15).
}
\label{carbonkinji1-1}
\end{center}
\end{figure}

As the nanotube length is finite, the integrand of (\ref{eq:denryu}) is a highly-oscillatory
function of $x$ with period $\sim 3\pi t/N$.
So we consider the average conductance: 
$\bar G=\langle \hat J \rangle/V$ where 
$V$ is the source-drain voltage.
In Fig. \ref{carbonkinji1-1}, the average conductance for $eV/t\sim 0.03$
is shown as a function of $\epsilon_{0}$
which corresponds to the gate voltage. 
The structural parameters of the nanotube
are $(N,M)=(7000,15)$, which corresponds to the length of $1.72$ $\mu$m and the diameter of $1.90$ nm.         

One can see that the envelope forms an asymmetric 
resonance peak around at $\epsilon _{0}-\epsilon _{F} \approx -0.2t$.
The case of $\epsilon _{0}=\epsilon _{F}$ corresponds to the
half-filled SWNT band and the flat region observed
from $\epsilon _{0}=\epsilon_{F}$ to $\epsilon _{0}=\epsilon_{F}-0.17t$
comes from the conduction band.
The sharp peak below $\epsilon _{0}=\epsilon_{F}-0.17t$ seems to 
arise from the van Hove singularity of the valence band
nearest to the conduction band.

We remark that the conductance is suppressed just below the conductance maximum 
(($\epsilon _{0}-\epsilon _{F})/t < -0.2$).
The dip seems to be caused by the interference between the two bands, the conduction
and valence bands.  
This means that the valence band nearest to the conduction band contributes to
the interference. 
In the system of the wave-guide geometry, even without the confined states,
similar suppression of the transmission due to the interference
has been reported \cite{Reichl}. 

We note that such asymmetric peaks are very sensitive to 
the coupling between the nanotube and electrodes.
In the above calculation, four tunneling matrix elements 
$\sum_{m=0}^{M-1}U_{m\lambda}^{(i)}e^{-i2a_{l'}m} (\lambda=L,R; i=1,2)$ 
involved in $h_{ll'}^{(\pm,\lambda)}$
are taken as the same real numbers.
This assumption implies that the nanotube 
contacts with both electrodes {\it via} single sites. 
This choice breaks the axial symmetry of the coupling, but    
the left and right contacts are equivalent.  
Indeed, when nanotube-electrode coupling is axially symmetric,
Fano-like shapes disappear in the conductance.
This could be understood as follows: Suppose that an asymmetric peak
arises from the interference between axially symmetric and asymmetric states.
Then, if the coupling is axially symmetric, non-symmetric nanotube
states do not contribute to the transport and the asymmetric peak
would disappear. \\   

\section{Conclusion}
We found that conductance shows Fano-like line shapes
even in a single SWNT due to the interference between
conduction and valence bands.
These peaks are strongly affected by
the contact condition. 

Our analysis implies that, even for the same SWNTs, Fano resonances
may or may not appear depending on the contact condition
with electrodes.
Therefore, we think that the observations by Kim {\it et. al. } \cite{Kim}
and by the Babi\'{c} \cite{Babic} are consistent.  

The exact form of the peaks depends on the nanotube structure 
and the contact conditions. Therefore, in order to reproduce experimentally
observed conductance, more detail information on the experimental
setting is required. Moreover, the neglected effects of the electron-electron
interaction should be taken into account.
In the present work, we have shown the possibility of Fano like conductance
peak and its dependence on the contact condition.
We believe the essence of these features is not altered.  
\section*{Acknowledgment}
The authors thank Professor M. Eto, Professor S. W. Kim,
Professor B. Li and Professor S. Pascazio 
for fruitful discussions and valuable comments. This work is 
partially supported 
by a Grant-in-Aid for Scientific 
Research of Priority Areas ``Control of Molecules in Intense Laser Fields'' 
and the 21st Century COE Program at Waseda University ``Holistic Research and 
Education Center for Physics of Self-organization Systems''
both from the Ministry of Education, Culture, Sports, Science and 
Technology of Japan 
and by Waseda University Grant for Special Research Projects
(2004A-161). 

\end{document}